# Analysing the use of graphs to represent the results of Systematic Reviews in Software Engineering

Katia Romero Felizardo
*Department of Computer Science*
*University of São Paulo – USP*
*São Carlos, Brazil*
*katiarf@icmc.usp.br*

Mehwish Riaz
*Department of Computer Science*
*University of Auckland*
*Auckland, New Zealand*
*mria007@aucklanduni.ac.nz*

Muhammad Sulayman
*Department of Computer Science*
*University of Auckland*
*Auckland, New Zealand*
*msul028@aucklanduni.ac.nz*

Emília Mendes
*Department of Computer Science*
*University of Auckland*
*Auckland, New Zealand*
*e.mendes@auckland.ac.nz*

Stephen G. MacDonell
*SERL, Comp. & Math. Sciences*
*AUT University*
*Auckland, New Zealand*
*stephen.macdonell@aut.ac.nz*

José Carlos Maldonado
*Department of Computer Science*
*University of São Paulo – USP*
*São Carlos, Brazil*
*jcmaldon@icmc.usp.br*

## Abstract

*The presentation of results from Systematic Literature Reviews (SLRs) is generally done using tables. Prior research suggests that results summarized in tables are often difficult for readers to understand. One alternative to improve results' comprehensibility is to use graphical representations. The aim of this work is twofold: first, to investigate whether graph representations result is better comprehensibility than tables when presenting SLR results; second, to investigate whether interpretation using graphs impacts on performance, as measured by the time consumed to analyse and understand the data. We selected an SLR published in the literature and used two different formats to represent its results - tables and graphs, in three different combinations: (i) table format only; (ii) graph format only; and (iii) a mixture of tables and graphs. We conducted an experiment that compared the performance and capability of experts in SLR, as well as doctoral and masters students, in analysing and understanding the results of the SLR, as presented in one of the three different forms. We were interested in examining whether there is difference between the performance of participants using tables and graphs. The graphical representation of SLR data led to a reduction in the time taken for its analysis, without any loss in data comprehensibility. For our sample the analysis of graphical data proved to be faster than the analysis of tabular data. However, we found no evidence of a difference in comprehensibility whether using tables, graphical format or a combination. Overall we argue that graphs are a suitable alternative to tables when it comes to representing the results of an SLR.*

## I. INTRODUCTION

The Systematic Literature Review (SLR) is a structured, well-organized, and step-by-step comprehensive method of conducting a review of the body of literature relevant to a particular research question. SLRs are often useful in identifying literature and research gaps relevant to a topic of interest [1]. The three main phases of an SLR include planning, conducting and reporting the review [1].

The results reported in an SLR are based on the data synthesized from the relevant identified literature. In this phase, the results of the primary studies – empirical studies investigating a specific research question [1] that meet the systematic review inclusion criteria - are summarized. This synthesis can be descriptive, but a quantitative summary obtained through the use of statistics can complement and lend greater weight to the description [2]. Biolchini et al. [3] suggest that the results obtained from an SLR must be displayed in tables to facilitate analysis, and that tables allow studies to be classified according to different criteria and to be organized under different perspectives. However, according to Kitchenham et al. [4] graphical representations of results are often easier for readers to understand than complicated tables.

Cruzes and Dybå [5] highlight that generally an SLR´s findings are presented in large tables, which contain a lot of data from individual studies (e.g., title, authors, year, outline, strengths, among others). However, a more useful tabular synthesis, such as a table combining findings, needs to be produced in the SLR context. Alternatively, the use of other visual representations could help to make the outcomes more readily understood.

Drawing on prior research, it is our contention that graphical representations should increase the comprehensibility of SLR outcomes and decrease the time required to analyse those outcomes. Thus, the main objective of this paper is to investigate the usefulness of graphs in comparison to tabular and mixed representations (i.e., both graphs and tables), in terms of comprehensibility and performance when used to represent the results of SLRs. The specific contribution of our work to the body of knowledge in the SLR field is to add empirical evidence regarding the effects (i.e., comprehensibility and performance) of the use of tabular/graphical representation to show outcomes of SLRs. These results should help reviewers when they decide to use tables, graphs or both representations together to show their findings.

The remainder of this paper is organized as follows. In Section 2 related work is presented. Section 3 presents the conducted experiment. Section 4 describes the results. The discussion, limitations, and future directions are presented in Section 5. Finally, Section 6 presents our conclusions.

## II. RELATED WORK

Edward Tufte, who developed foundational theories about the visual display of information, stated that graphical excellence consists of complex ideas communicated with clarity, precision, and efficiency [12]. Humans have strong visual processing abilities and visual representations can be exploited by the human system to support knowledge discovery; thus the central aim of visualization is to help users explore and understand data [6].

Visualization enables users to "see" and navigate through data in multiple ways, with their optical abilities enhancing the knowledge acquisition process [13]. The intent of visual data exploration is to visually represent relevant information and enable the user to interact with the information, gain insights and detect interesting knowledge [7]. Prior research indicates that information users have confidence in findings shown using visual representations, which are both more intuitive and more rapidly explored [6]. As a result, graphical representations of information may be preferable to the use of tabular or textual reports [7; 8]. Research in other fields has found that, compared with tables, the use of graphical representations to present empirical results increases the clarity of presentation and makes it easier for a reader to understand the data [9].

Outside the specific context of SLRs the 'graphs vs. tables' question has been addressed for some time. Jarvenpaa and Dickson [10] affirm that in specific cases - to summarize data (which is one reason for undertaking an SLR [1]); to show trends and relationships over time; to compare data points and relationships of variables; and to detect deviations or differences in data - the use of graphics is more indicated than tables. Vessey [11] remarks, however, that despite the extent of research that has compared decision-making performance when using graphical and tabular representations, there are few conclusions about the thematic. More specifically, there is a lack of research that has investigated the application of graphical representations in the context of SLRs, especially in the reporting phase in which there is often a need for data synthesis [5].

In the context of Evidence Based Software Engineering (EBSE), Garcia et al. [14] analysed how graphical representations, such as parallel coordinates, may complement statistical data analysis, helping users to understand and treat data from empirical studies. This research was the first initiative towards introducing graphical representations in the analysis of data from empirical studies in SE; however the data analysed were of one experiment replication conducted in a specific scope (i.e., the application of several reading techniques, aimed at evaluating and comparing their efficacy and efficiency).

Cruzes and Dybå [5] performed a tertiary review to assess the types and methods of research synthesis evident in SLRs in Software Engineering (SE). They included 31 studies in their review and found that almost half of those studies (13 of the 31 considered) did not contain any synthesis. This suggests that currently the attention given to research synthesis in SE may be limited. The authors reported that just over half of the studies analysed used tables (i.e., the simplest type of graphic presentation) to show the findings. Other forms of visual representation were used in fewer than 20% of the studies. The authors mention that, beyond tables, other visual representations can be used to represent the results of an SLR. One example is the graph, which can show complex findings, displaying the connections among the primary studies. They go on to note that when findings are complex when represented in a table format, graphs are a useful choice to visualize those results. One example of the use of graphs in an SLR can be found in [15]. This study evaluated relations between the regression test selection techniques and its results were shown in graphs.

Malheiros et al. [16] investigated the use of visual text mining (VTM) techniques to assist the study selection activity in the SLR process. In their study they used a VTM document map as a visual representation of the SLR's primary studies.

## III. EXPERIMENT

The aim of the work presented here is twofold: first, to investigate whether graphical representations, such as graphs, deliver better comprehensibility than tables when presenting SLR results (i.e., enhancing a reader's ability to understand the data); second, to investigate whether data analysis using graphs impacts on its performance (i.e., the time consumed to analyse and understand the data). In this study, we assert that the use of graphs may positively influence the comprehensibility of the SLR's results. Furthermore, we believe that the use of graphs to represent the results of SLRs will positively affect reader performance (amount of time required) in comprehending data.

For the purpose of this research, we are interested in a particular type of graphical representation, i.e., graphs. A graph is an abstract data structure that consists of a finite set of ordered pairs, called edges and nodes [17]. The nodes represent objects (that can be tangible or intangible depending on the application) and these are connected by edges that can refer to some common shared aspects. In the context of SLRs, the graph structure – nodes and edges – can reveal, for example, authors, primary studies (an individual piece of evidence, i.e., a case study or an experimental study), and the relationships between them.

There are many reasons why graphs are considered to be powerful visualization tools: (i) graphs are structurally simple models (comprising nodes and edges) that can be applied to various applications, such as social networks and paper citations; (ii) graphs can usefully represent data and the relations among them – an example is the Internet, in which the nodes could be web pages and the edges could represent hyperlinks (relationships); (iii) comparing graphs with other visual representations (e.g., pie graphs and box plots), graphs permit better representation of data with internal relationships [18]; (iv) even though graphs are an abstract concept, the graph theory field has a very solid foundation and there are several algorithms for processing graphs efficiently [19; 20]. As a result, there is extensive interest in graph theory and its many areas of application, such as computer vision and image processing, robotics, network analysis, web mining, chemistry, bioinformatics, sensor networks, biomedical engineering and evolutionary computation [21].

As stated, graphs can be applied to visualize any kind of data where there is an inherent internal relation among the data elements [22]. In table visualizations the data are generally represented in a two-dimensional structure (i.e., rows, which represent objects; and columns, which represent variables or dimensions). There is no explicit mechanism to depict hierarchical or network structures i.e., links among the data [23].

In order to investigate the usefulness of graphs in representing the results of SLRs, an experiment was conducted where forms (in a questionnaire format) were created using tables, graphs, and a combination of tables and graphs to represent the outcomes of an SLR. Our study may be characterized as follows:

- Object of study: Graphs.
- Purpose: To improve the analysis results of SLRs.
- Focus: Comprehensibility and performance.
- Perspective: From the point of view of researchers.
- Context: The representation of SLR results in the field of SE and their interpretation by researchers, including Master's and PhD students and experts in SLRs.

The following subsections detail the experimental design including the preparation of the experimental instrument and data collection.

### *A. Preparation of the Experimental Instrument*

The experimental instrument was carefully created using as source an existing SLR, by converting the SLR results into a graphical representation, and creating the relevant forms, each discussed in the following subsections.

#### *Source SLR for the Experimental Instrument*

In order to evaluate the comprehensibility of results using graphs, we chose the SLR entitled "Systematic literature reviews in software engineering – A systematic literature review" [24] for three main reasons: (1) it was conducted by Kitchenham, a widely acknowledged SLR expert who defined the guidelines for performing SLRs in SE [1]; (2) the SLR's results were published in table format and; (3) the topic addressed is general, it describes the impact of SLRs in SE based on the following research questions (RQs) [24]:

- RQ1: How many SLRs were published between 1st January 2004 and 30th June 2008?
- RQ2: What research topics are addressed?
- RQ3: Which individuals and organizations are most active in SLR-based research? and
- RQ4: Is the quality of SLRs improving?

#### *Conversion of tables into graphs*

To convert the source tables into graphs, one of the authors reproduced the results originally shown in table format in a graphical format. The information contained in the set of tables described in Kitchenham et al. [24] was restructured to be used in an open source tool called Projection Explorer (PEx) [25]. The PEx tool was originally built to realize the projection of multidimensional collections of text documents and was later adapted to display collections of images, temporal series, among others. This work uses an extension of the tool for graphs – PEx-Graph [26] – that implements specific features such as to enable visual attributes of the nodes to be changed, e.g. changing colour to represent the type of the object.

Figure 1 shows an example of the conversion from a table to a graph. The table-graph conversion process is executed in two steps:

- Initially, it is necessary to re-write the information presented in a table (Figure 1a) in PEx-Graph input format (Figure 1b). At the top of the document we have the set of information about the objects (nodes) and at the bottom we have the relationships (edges);
- Next, the nodes are drawn as circles, and nodes that have relationships between them are connected by lines representing the edges. An example of a final visualization is shown in Figure 1c. As mentioned, the graph nodes can represent more than one type of object; in this example, they represent paper, year and country (location of the authors). Moreover, it is possible to observe that the edges connect nodes of different types, such as relationships between a paper and its corresponding year of publication and country of origin.

Note that we have deliberately chosen a simple example here so that the basic idea underpinning the graphical representation of SLR outcomes can be easily understood – the real benefits from using such a visualization become more evident in representing more complex data sets, as shown in later sections of the paper.

The table-graph conversion process was applied to create graphs related to the four RQs addressed by Kitchenham et al. [24]. Table 1 represents an extract of the original table presented by Kitchenham et al. [24] containing information related to RQ1 (*How many SLRs were published between 1st January 2004 and 30th June 2008?*). Figure 2 shows the graph related to RQ1, after the application of the table-graph conversion process. The figure contains exactly the same information as is presented in Table 1 and the graph is the direct result of processing by the tool i.e. no user actions were taken to adjust or improve the visualization.

*Forms Preparation*

The forms used in our experiment contained several questions that were chosen based on four data sets, one data set related to each of the four RQs from Kitchenham et al. [24]. Three different forms were created, each containing the same results but represented in three different formats: (i) table format; (ii) graph format and (iii) a combination of both tables and graphs. In relation to the form that contained both formats, the first and third data sets were represented in graph format and the second and fourth were represented in tabular format.

| Paper | Year | Country |
|---|---|---|
| *Title of paper 1* | *2009* | *country_1* |
| *Title of paper 2* | *2010* | *country_2* |
| *Title of paper 3* | *2009* | *country_1* |

***(a)*** *Table representation*

****Node data***
***ID, type, name***
*paper_1, paper, "Title of paper 1"*
*paper_2, paper, "Title of paper 2"*
*paper_3, paper, "Title of paper 3"*
*country_1, country, "Name of country 1"*
*country_2, country, "Name of country 2"*
*year_1, year, "2009"*
*year_2, year, "2010"*

**Nodes' information** (paper, year, country)

****Tie data***
***from, to, relationship***
*paper_1, country_1, 1*
*paper_2, country_2, 1*
*paper_3, country_1, 1*

**Edges' Information**

*paper_1, year_1, 1*
*paper_2, year_2, 1*
*paper_3, year_1, 1*

**Edges' Information**

***(b)*** *PEx-Graph: input representation*

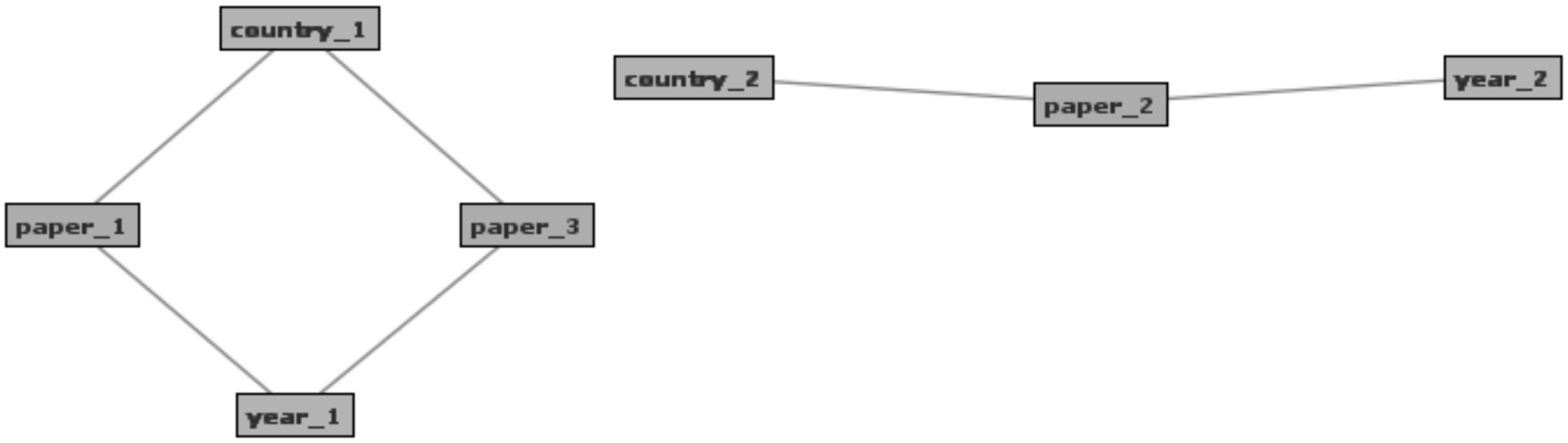

***(c)*** *PEx-Graph: output representation*

**Fig. 1.** Conversion activity to create visual graphs using PEx-Graph tool.

**TABLE 1**

Conversion Information related to RQ1 [24]: Primary Study and its respective type and year.

| Primary Study | Type | Year |
|---|---|---|
| p_1 | MS | 2006 |
| p_2 | MS | 2007 |
| p_3 | MS | 2007 |
| p_4 | SLR | 2004 |
| p_5 | MS | 2007 |
| p_6 | SLR | 2005 |
| p_7 | SLR | 2005 |
| p_8 | MS | 2008 |
| p_9 | SLR | 2006 |
| p_10 | SLR | 2007 |
| p_11 | SLR | 2007 |
| p_12 | SLR | 2006 |
| p_13 | SLR | 2007 |
| p_14 | SLR | 2005 |
| p_15 | SLR | 2004 |
| p_16 | SLR | 2004 |
| p_17 | MS | 2008 |
| p_18 | SLR | 2007 |

| Primary Study | Type | Year |
|---|---|---|
| p_19 | SLR | 2008 |
| p_20 | MS | 2007 |
| p_21 | MA | 2005 |
| p_22 | SLR | 2006 |
| p_23 | SLR | 2007 |
| p_24 | MS | 2007 |
| p_25 | SLR | 2005 |
| p_26 | SLR | 2005 |
| p_27 | SLR | 2008 |
| p_28 | MS | 2008 |
| p_29 | SLR | 2008 |
| p_30 | MS | 2008 |
| p_31 | MS | 2005 |
| p_32 | SLR | 2006 |
| p_33 | SLR | 2006 |
| p_34 | SLR | 2004 |
| p_35 | SLR | 2007 |
| p_36 | SLR | 2008 |

| Primary Study | Type | Year |
|---|---|---|
| p_37 | SLR | 2004 |
| p_38 | SLR | 2004 |
| p_39 | SLR | 2006 |
| p_40 | SLR | 2008 |
| p_41 | MS | 2007 |
| p_42 | MS | 2007 |
| p_43 | SLR | 2008 |
| p_44 | SLR | 2005 |
| p_45 | SLR | 2007 |
| p_46 | MS | 2007 |
| p_47 | SLR | 2005 |
| p_48 | MS | 2005 |
| p_49 | SLR | 2006 |
| p_50 | MS | 2008 |
| p_51 | MS | 2005 |
| p_52 | SLR | 2006 |
| p_53 | SLR | 2008 |

**Legend:**
Systematic Literature Review (SLR)
Meta-Analysis (MA)
Mapping studies (MS)

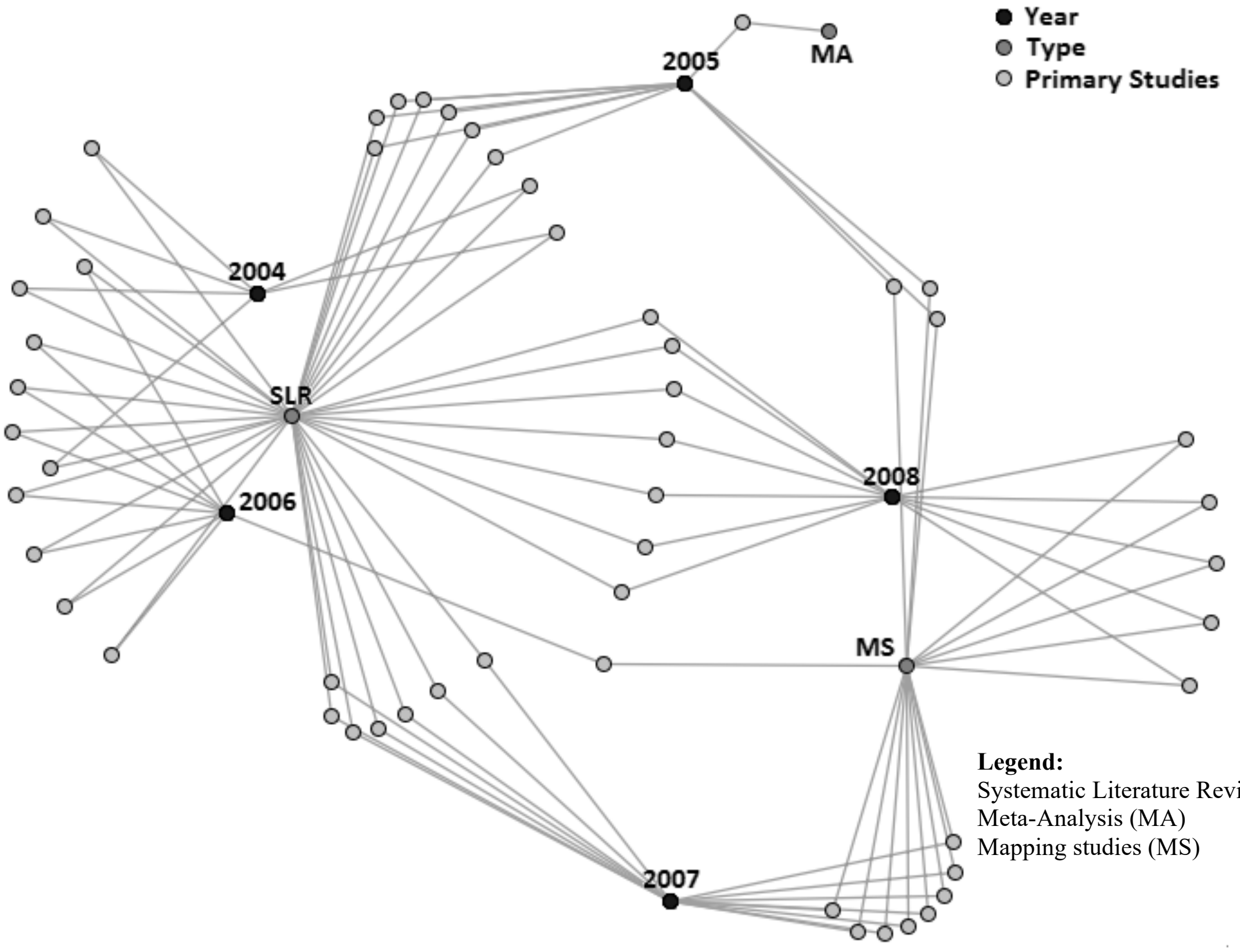

**Legend:**
Systematic Literature Review (SLR)
Meta-Analysis (MA)
Mapping studies (MS)

**Fig. 2.** Graph relative to RQ1 [24].

The results of the source SLR [24] were presented in such a way that they could easily be converted into graphical format. For example, analysing Table 1 or Figure 2, it is possible to observe (as mentioned by Kitchenham et al.) that in 2004 – the year in which the SLR was proposed to be used in SE as a method for aggregating evidence – is the year with the lowest number of published papers, at only 6. Furthermore, 2007 was the year with the highest number of papers, 15 in total. It can also be concluded that in 2008, there were 12 published papers. The 53 primary studies can also be grouped by their types: 35 papers were based on Systematic Literature Reviews (SLR); 1 was based on Meta-Analysis (MA); and 17 were based on Mapping Studies (MS). Thus, it is possible to affirm that SLRs were the most frequently occurring type of evidence-based study and that meta-analysis, common in other disciplines, was very rare in SE. Analysing the combination of the type of publications through the years, in the year 2004 all papers were SLRs. It is possible to come to the same conclusions using either the information contained in Table 1 or in Figure 2 (the graph). Thus, the questions we designed for the experimental instrument related to RQ1 and these data sets were:

- In which year the least number of papers was published?
- In which year the highest number of papers was published?
- In which year 12 papers were published?
- The highest number of papers published was published of type "SLR? Yes ( ) No ( )
- In which year all published papers were of type "SLR"?

The same process was used to design the questions related to the other 3 data sets. In this way we compiled a total of 13 questions: 5 related to RQ1 (above), 4 related to RQ2, 2 related to RQ3 and 2 related to RQ4.

In order to validate the contents of the experimental instrument i.e., the forms, the experiment was pilot run where one of the authors of this paper (MacDonell), an expert in conducting SLRs, participated. Note that the data gathered through the pilot run of the experiment was discarded and not considered in our findings.

After the forms were validated, the final version of the forms contained a task list, comprising four tables and/or graphs to be analysed and their respective questions. The forms can be downloaded from the following web location: *http://www.labes.icmc.usp.br/katia/graphs/forms.rar*

### *B. Data Collection*

With the forms validated, requests for participation were sent via email to researchers, PhD and Masters students, in SE departments of two universities in New Zealand: (1) University of Auckland and (2) Auckland University of Technology. We received responses from 7 PhD students (i.e., 2 using tables, 4 using graphs and 1 using tables and graphs combined i.e. a form that included a mix of some graphs and some tables) and 4 experts in SLRs (i.e., 1 using tables, 2 using graphs and 1 using tables and graphs combined). To get similar size data sets for each of the three forms in order to validate the comparisons, requests for participation were sent via email to other researchers, PhD and Masters students, in the SE department at the University of São Paulo (USP), Brazil. As a result, over the whole experiment, data were collected from 6 experts in conducting SLRs (i.e., 2 using each of the three forms); 12 PhD students (i.e., 4 using each of the three forms); and 6 Masters students (i.e., 2 using each of the three forms).

## IV. RESULTS

The results presented in this section are based on the data gathered from 24 participants with varying levels of experience with research and SLRs – eight participants for each type of form i.e., graphs, tables, and a combination of graphs and tables. To avoid a language effect we ensured that all Brazilian participants had English language proficiency. Furthermore, of all the Master's student respondents only 1 participant (number 5) did not have prior experience in conducting SLRs.

Table 2 presents a summary of the results. This table is split into three sections – one for each type of form. For each type of form, we show the comprehension score obtained by each participant (out of 13), the percentage of the scores obtained by each participant, the self-reported time taken by each participant to complete the form, and finally the average percentage score and average time taken.

The results of the experiment show that the average percentage scores for graphs, tables, and a combination of graphs and tables are 93.3%, 89.4%, and 86.5%, respectively. This suggests that the comprehensibility of the graph representation was greater than that of the tables and the tables and graphs combined. However, we applied ANOVA testing to evaluate whether the differences among these results were caused by chance or were legitimate (see Table 3 – Score Average) [27; 28] and this showed that the difference between the average scores of the participants using the three forms was not statistically significant.

One explanation for such a finding is advocated by Vessey [11], who affirmed that graphs and tables support different types of task, i.e., graphs support spatial tasks, which require the user to make associations and to perceive relationships in the data, whereas tables support symbolic tasks, which involve the extraction of discrete data values.

To better understand our findings, one of the authors classified the 13 tasks of our forms into the two classes defined by Vessey [11] (see Table 5). Therefore, we had 4 tasks classified as symbolic (tasks 3, 5, 7 and 13) and 9 as spatial (tasks 1, 2, 4, 6, 8, 9, 10, 11 and 12). For validation purposes, the same tasks were independently classified by another researcher and the level of agreement between the first and second researchers was 100%. Using this binary classification, we analysed whether the comprehensibility for spatial tasks using graphs was higher than that of the tables, and the inverse situation, i.e., whether the comprehensibility for symbolic tasks using tables was higher than that of the graphs (see Tables 6 and 7).

**TABLE 2**
Results Summary.

| No. | Experience | Score | % Score | Time |
|---|---|---|---|---|
| **Graphs** | | | | |
| **1** | Master 1 | 12/13 | 92.3 | 12' 00" |
| **2** | Master 2 | 12/13 | 92.3 | 07' 00" |
| **3** | PhD 1 | 12/13 | 92.3 | 10' 23" |
| **4** | PhD 2 | 13/13 | 100 | 11' 00" |
| **5** | PhD 3 | 12/13 | 92.3 | 11' 00" |
| **6** | PhD 4 | 12/13 | 92.3 | 12' 00" |
| **7** | Expert 1 | 13/13 | 100 | 05' 00" |
| **8** | Expert 2 | 11/13 | 84.62 | 04' 00" |
| | | Average: | 93.3 | 09' 15" |
| **Tables** | | | | |
| **9** | Master 3 | 11/13 | 84.62 | 52' 00" |
| **10** | Master 4 | 12/13 | 92.3 | 45' 00" |
| **11** | PhD 5 | 11/13 | 84.62 | 13' 50" |
| **12** | PhD 6 | 13/13 | 100 | 47' 00" |
| **13** | PhD 7 | 11/13 | 84.62 | 34' 00" |
| **14** | PhD 8 | 12/13 | 92.3 | 21' 00" |
| **15** | Expert 3 | 12/13 | 92.3 | 31' 00" |
| **16** | Expert 4 | 11/13 | 84.62 | 09' 00" |
| | | Average: | 89.4 | 31' 56" |
| **Mixed (Graphs and Tables)** | | | | |
| **17** | Master 5 | 12/13 | 92.3 | 22' 00" |
| **18** | Master 6 | 11/13 | 84.62 | 26' 00" |
| **19** | PhD 9 | 11/13 | 84.62 | 10' 00" |
| **20** | PhD 10 | 12/13 | 92.3 | 08' 00" |
| **21** | PhD 11 | 12/13 | 92.3 | 17' 00" |
| **22** | PhD 11 | 11/13 | 84.62 | 22' 00" |
| **23** | Expert 5 | 08/13 | 61.54 | 11' 00" |
| **24** | Expert 6 | 13/13 | 100 | 22' 00" |
| | | Average: | 86.5 | 17' 25" |

The ANOVA testing (see Table 3) confirms our previous results, showing that the difference between the score averages of the participants in each of the two tasks using the different formats (tables and graphs) is not significant. Thus, we conclude that the accuracy in comprehending results presented in tables is no different to that of graphs.

The results of the experiment also reveal that the average time taken to understand the results presented in the form of graphs, tables, and graphs and tables combined, is 9'15", 31'56", and 17'25", respectively (see Table 2). This shows that the graphs were most efficiently understood by the participants, followed by the combination of graphs and tables, and finally tables only. The data presented in the form of tables took on average the largest amount of time for the participants to understand. ANOVA testing (see Table 3 – Time) confirms that the difference between the average time taken by the participants to understand the data presented in each of the three forms is significant.

To test all pairwise comparisons among time means, we used the Tukey test [27; 28] (see Table 4). The results show that there is a statistically significant difference between the time averages for the use of tables and graphs. There is, also, statistical significance between the time averages of tables and the combined approach. However, there is no statistical significance for the difference between the time averages of graphs and the combination of graphs and tables (mix). Thus, graphs, overall, have been observed as the least time-consuming format for understanding the data presented in the SLR. On the other hand, tables have been observed as the higher time-consuming format.

**TABLE 3**
Summary of results for ANOVA testing.

| Dataset | Variable Compared | P-Value | Statistically Significant? |
|---|---|---|---|
| Graphs, Tables, Mix | Score Average | 0.2561 | No (P-Value > 0.05) |
| | Time | 0.00090 | Yes (P-Value < 0.05) |

**TABLE 4**
Summary of results for Tukey testing.

| Dataset | Variable Compared | P-Value | Statistically Significant? |
|---|---|---|---|
| Graphs, Tables | Time | 0.000644651 | Yes (P-Value < 0.05) |
| Graphs, Mix | Time | 0.262373897 | No (P-Value > 0.05) |
| Tables, Mix | Time | 0.026441964 | Yes (P-Value < 0.05) |

**TABLE 5**
Classification of the questions of our questionnaire according to Vessey [11].

| Task | Question | Classification | Justification |
|---|---|---|---|
| 1 | 1. In which year the least number of papers was published? | Spatial | This question requires a comparison of all data values to define the least value. |
| | 2. In which year the highest number of papers was published? | Spatial | This question requires a comparison of all data values to define the highest value. |
| | 3. In which year 12 papers were published? | Symbolic | This question requires a specific year as the response. |
| | 4. The highest number of papers published was published of type “SLR”? | Spatial | This question requires a comparison of all data values to define the highest value. |
| | 5. In which year all published papers were of type “SLR”? | Symbolic | This question requires a specific year as the response. |
| 2 | 6. How many different topics have been addressed? | Spatial | This question requires a comparison of all topics to define whether they are different. |
| | 7. How many studies have investigated Software Architecture? | Symbolic | This question requires a specific number as the response. |
| | 8. Which topic has been investigated most often? | Spatial | This question requires a comparison of all data values to define the most often. |
| | 9. There have been more studies on Software Process Improvement than on Tool Integration. | Spatial | This question requires a comparison between both topics, Software Process Improvement and Tool Integration. |
| 3 | 10. Which country has produced the highest number of published SLRs? | Spatial | This question requires a comparison of all data values to define the highest value. |
| | 11. Which country which has produced the second highest number of SLRs? | Spatial | This question requires a comparison of all data values to define the second highest value. |
| 4 | 12. Do you think that the guidelines influence the quality score? | Spatial | This question requires assessing relationship (score x guidelines) in the data. |
| | 13. How many studies scored 2.5? | Symbolic | This question requires a specific number as the response. |

**TABLE 6**
Summary of results for the Symbolic Tasks.

| | Symbolic Task | Master 1 | Master 2 | PhD 1 | PhD 2 | PhD 3 | PhD 4 | Expert 1 | Expert 2 | Correct answers ✓ |
|---|---|---|---|---|---|---|---|---|---|---|
| **Using Tables** | **3** | ✓ | ✓ | ✓ | ✓ | ✓ | ✓ | ✓ | ✓ | **8/8** |
| | **5** | ✓ | ✗ | ✓ | ✓ | ✓ | ✓ | ✓ | ✓ | **7/8** |
| | **7** | ✓ | ✓ | ✓ | ✓ | ✓ | ✓ | ✓ | ✓ | **8/8** |
| | **13** | ✓ | ✓ | ✓ | ✓ | ✓ | ✓ | ✓ | ✓ | **8/8** |
| **Using Graphs** | **Symbolic Task** | **Master 3** | **Master 4** | **PhD 5** | **PhD 6** | **PhD 7** | **PhD 8** | **Expert 3** | **Expert 4** | **Correct answers ✓** |
| | **3** | ✓ | ✓ | ✓ | ✓ | ✓ | ✓ | ✓ | ✓ | **8/8** |
| | **5** | ✓ | ✓ | ✓ | ✓ | ✓ | ✗ | ✓ | ✓ | **7/8** |
| | **7** | ✓ | ✓ | ✓ | ✓ | ✓ | ✓ | ✓ | ✗ | **7/8** |
| | **13** | ✗ | ✗ | ✓ | ✓ | ✗ | ✓ | ✓ | ✓ | **5/8** |

**TABLE 7**
Summary of results for the Spatial Tasks.

| | Spatial Task | Master 1 | Master 2 | PhD 1 | PhD 2 | PhD 3 | PhD 4 | Expert 1 | Expert 2 | Correct answers ✓ |
|---|---|---|---|---|---|---|---|---|---|---|
| **Using Tables** | **1** | ✓ | ✓ | ✓ | ✓ | ✓ | ✓ | ✓ | ✓ | **8/8** |
| | **2** | ✗ | ✓ | ✓ | ✓ | ✗ | ✓ | ✓ | ✓ | **6/8** |
| | **4** | ✓ | ✓ | ✓ | ✓ | ✓ | ✓ | ✓ | ✓ | **8/8** |
| | **6** | ✗ | ✓ | ✗ | ✓ | ✗ | ✗ | ✗ | ✗ | **2/8** |
| | **8** | ✓ | ✓ | ✓ | ✓ | ✓ | ✓ | ✓ | ✓ | **8/8** |
| | **9** | ✓ | ✓ | ✓ | ✓ | ✓ | ✓ | ✓ | ✗ | **7/8** |
| | **10** | ✓ | ✓ | ✓ | ✓ | ✓ | ✓ | ✓ | ✓ | **8/8** |
| | **11** | ✓ | ✓ | ✓ | ✓ | ✓ | ✓ | ✓ | ✓ | **8/8** |
| | **12** | ✓ | ✓ | ✗ | ✓ | ✓ | ✓ | ✓ | ✓ | **7/8** |
| **Using Graphs** | **Spatial Task** | **Master 3** | **Master 4** | **PhD 5** | **PhD 6** | **PhD 7** | **PhD 8** | **Expert 3** | **Expert 4** | **Correct answers ✓** |
| | **1** | ✓ | ✓ | ✓ | ✓ | ✓ | ✓ | ✓ | ✓ | **8/8** |
| | **2** | ✓ | ✓ | ✓ | ✓ | ✓ | ✓ | ✓ | ✓ | **8/8** |
| | **4** | ✓ | ✓ | ✓ | ✓ | ✓ | ✓ | ✓ | ✓ | **8/8** |
| | **6** | ✓ | ✓ | ✓ | ✓ | ✓ | ✓ | ✓ | ✗ | **7/8** |
| | **8** | ✓ | ✓ | ✓ | ✓ | ✓ | ✓ | ✓ | ✓ | **8/8** |
| | **9** | ✓ | ✓ | ✗ | ✓ | ✓ | ✓ | ✓ | ✓ | **7/8** |
| | **10** | ✓ | ✓ | ✓ | ✓ | ✓ | ✓ | ✓ | ✓ | **8/8** |
| | **11** | ✓ | ✓ | ✓ | ✓ | ✓ | ✓ | ✓ | ✓ | **8/8** |
| | **12** | ✓ | ✓ | ✓ | ✓ | ✓ | ✓ | ✓ | ✓ | **8/8** |

**TABLE 8**
Summary of results for ANOVA testing
(Symbolic and Spatial tasks).

| Dataset | Variable Compared | P-Value | Statistically Significant? |
|---|---|---|---|
| Graphs, Tables | Score Average (Symbolic tasks) | 0.1901 | No (P-Value > 0.05) |
| Graphs, Tables | Score Average (Spatial tasks) | 0.2041 | No (P-Value > 0.05) |

The results obtained for participant 5, who did not have prior experience in conducting SLRs, were similar to those of other participants with the same level of experience (PhD). For instance, participant 5 took 11'00" to analyse the graphs, and participants 3, 4 and 6 took 10'23", 11'00" and 12'00", respectively. The percentage score of participant 5 was 92.3%, and the percentage scores of participants 3, 4 and 6 were 92.3%, 100% and 92.3%, respectively. As a result we do not believe that participant 5's results were inconsistent with those of participants with similar levels of prior experience.

## V. DISCUSSION

This paper investigates the use of graphs to present the results of SLRs conducted in the domain of SE. For this purpose, the results of an SLR [24] were converted to graphical representations and an experiment was conducted to assess the accuracy and efficiency in comprehending the results of the SLR using three different forms – graphs, tables, and a combination of graphs and tables.

Our results show that the results presented in the form of graphs are most efficiently understood in comparison to those presented in the form of tables, or tables and graphs combined. The results of the experiment also show that while there was a difference in the comprehensibility scores obtained by the participants for each type of form, these differences were not statistically significant. We would contend that the non-significant difference in the comprehensibility scores is due to the significant amount of time taken to answer the 13 questions. In other words, the comprehensibility scores will increase with an increase in the time taken to answer the questions. Therefore, based on our results, we believe that if equal amounts of time were allocated to each group of participants (i.e., for each type of form), the comprehensibility scores of the group analyzing graphs would have been much higher than those of the other two forms i.e., tables, and graphs and tables combined.

Based on our results, we conclude that graphs are a useful, and in some cases better, representational format in which to present the results of SLRs. The use of graphs increases the comprehensibility of the presented results when taking into account accuracy of understanding and time taken to understand the results. Therefore, we suggest that graphs should be used more extensively, whenever applicable, to present the results of SLRs conducted in the domain of SE. This recommendation takes into account the additional workload imposed by the table to graph conversion – utilizing the PEx-Graph takes only some seconds in order to create and present the visual representations.

For the purpose of this study, the data from the source SLR was converted to only one particular type of graphical representation, i.e., a set of nodes and edges. Other types of graphical diagrams such as bar charts, pie charts, line charts, scatterplots or boxplots may be useful depending upon the type of data to be presented.

This study, like most research, comes with certain limitations. The first limitation is that the results are based on only one source SLR where the form of the presented results could easily be converted to graphs. Since the results of SLRs are not always in a format that can easily be converted to a graphical representation and in many cases there is a need to present textual data, the results may not be presented fully if graphs are used. Therefore, it may not always be possible to obtain a higher level of accuracy and efficiency in comprehending the results of the SLRs presented in the form of graphs in comparison to tabulated or textual representations. The second limitation relates to the external validity of the research since our results are based on only 8 respondents for each of the three types of form. However, for a first experiment of this nature, we would contend that a total number of 24 participants with varying levels of experience in research and conducting SLRs is a usefully indicative sample. Future work could involve carrying out similar experiments with larger sample sizes, considering other forms of visualization, for example, bar charts, pie charts, line charts, scatterplots or boxplots, and using results from other SLRs. A related limitation is that in our using one specific type of graph it is not possible to conclude that graphs are better in general.

## VI. CONCLUSION

This study investigated the impact on the level of comprehensibility, determined by two variables - accuracy and efficiency, of using graphical representations to present the results of SLRs. The results of a previously published SLR [24] presented in a tabular form were converted into graphical representation using PEx-Graph. Three types of forms were created with three representations i.e., tables, graphs, and a combination of graphs and tables, of the same data based on the results of the source SLR. An experiment was conducted with 24 participants with varying levels of experience with research and conducting SLRs where the time and accuracy to answer 13 questions listed in the forms were recorded. The results of the experiment show that graphs were most efficiently understood by the participants, i.e. there is a reduction in the time taken for their analysis.